\documentclass[preprints,article,accept,oneauthor,pdftex]{Definitions/mdpi} 

\usepackage{graphicx}
\usepackage{bm}
\usepackage{dsfont,amsmath,amssymb}
\usepackage[mathscr]{eucal}
\usepackage{amsthm}

\newcommand{\Tr}{\mathrm{Tr}}
\newcommand{\sgn}{\mathrm{sgn}}

\newcommand{\ee}{\mathrm{e}}
\newcommand{\ii}{\mathrm{i}}
\renewcommand{\SS}{\mathscr{S}}
\newcommand{\EE}{\mathscr{E}}
\newcommand{\SE}{\mathscr{SE}}
\newcommand{\HH}{\mathcal{H}}

\firstpage{1} 
\pubvolume{}
\issuenum{}
\articlenumber{}
\pubyear{}
\copyrightyear{}
\history{} 

\pdfoutput=1

\Title{Quantum Thermodynamics in the Refined Weak Coupling Limit}

\Author{\'Angel Rivas$^{1,2}$}

\AuthorNames{\'Angel Rivas}

\address{%
$^{1}$ \quad Departamento de F\'isica Te\'orica, Facultad de Ciencias F\'isicas, Universidad Complutense, 28040, Madrid, Spain.\\
$^{2}$ \quad CCS-Center for Computational Simulation, Campus de Montegancedo UPM, 28660 Boadilla del Monte, Madrid, Spain.}

\corres{anrivas@ucm.es}

\abstract{We present a thermodynamic framework for the refined weak coupling limit. In this limit the interaction between system and environment is weak, but not negligible. As a result, the system dynamics becomes non-Markovian breaking divisibility conditions. Nevertheless, we propose a derivation of the first and second law just in terms of the reduced system dynamics. To this end, we extend the refined weak coupling limit for allowing slowly-varying external drivings, and reconsider the definition of internal energy due to the non-negligible interaction.}


\begin{document}

\tableofcontents

\section{Introduction}
The theory of open quantum systems describes the evolution of a quantum system that exchanges energy or information with some environment \cite{BrPr02,GardinerZoller04,Libro}. The understanding on whether a thermodynamical framework can be applied to these processes is not only of fundamental, but also of practical interest and constitutes the active field of quantum thermodynamics \cite{Qthermo1,QthermoKosloff,Qthermo2}. In analogy with the classical case, one speaks about equilibrium quantum thermodynamics if the open quantum system considered is in equilibrium with its environment, analyzing the change of the system's thermodynamic properties when there is a change of one equilibrium state into another equilibrium state. On the contrary, nonequilibrium quantum thermodynamics focuses on the change of those properties during the evolution of the open quantum system due to the environmental interaction. Provided that the coupling between system and environment is sufficiently weak, this evolution can be approximated by the celebrated weak coupling limit \cite{BrPr02,Libro,Davies}, which allows for the formulation of a dynamical equation for the density matrix of the system with the Gorini-Kossakowski-Lindblad-Sudarshan (GKLS) form \cite{GKLS1,GKLS2}. In absence of external driving, this equation generates a quantum dynamical semigroup \cite{AlickiBook} with very appealing thermodynamical properties. Namely, the system approaches asymptotically the equilibrium state \cite{AlickiBook,BrPr02,Libro}, the entropy production is positive \cite{SpohnEntropy,QthermoKosloff}, the heat is additive under the presence of several thermal reservoirs in the environment \cite{SpohnLebowitz,QthermoKosloff}, etc. Moreover, work can be incorporated into this framework under some conditions, such as slowly-varying or periodically-varying drivings \cite{AlickiHeatEngine,AlickiTutorialPeriodic}. However, if the coupling between system and environment is not weak, the evolution of the open system becomes typically non-Markovian, and the formulation of a thermodynamic framework is more complicated \cite{Esposito,Modi,Tanimura,Alipour1,Alipour2,StrasbergPRX,Pati,Ghosh,StrasbergPRE}. This is so even for the equilibrium case \cite{Gelin,Seifert,Hu}.

It is the aim of this contribution to introduce a valid thermodynamics framework for the evolution obtained in a refined weak coupling limit \cite{AlickiRefined,SchallerBrandes,Benatti1,Refined,Libro}. This technique still considers the system-environment coupling to be weak, but not negligible at short times. As a consequence, the open system dynamics becomes non-Markovian, and it can actually present strong non-Markovian properties like positive indivisibility (P-indivisibility) or ``quasieternal'' completely positive indivisibility (CP-indivisibility) \cite{Refined}. We suggest a reformulation of the internal energy in this regime, such that the entropy production is never negative because the obtained dynamical map is, by construction, completely positive (CP). Moreover, in order to allow for the existence of work, we introduce an extension of the refined weak coupling technique for slowly-varying system Hamiltonians, and establish the first and second law in this setting. As a key difference with other approaches, our thermodynamic relations only involve system observables. Namely, they are completely formulated in terms of the system reduced dynamics. This is something natural in the Markovian case, but very challenging to be satisfied for non-Markovian evolution.

The present paper is structured as follows: in Sec. 2 we briefly explain the usual weak coupling limit and the refined one, indicating their principal differences. Sec. 3 is a succinct, but self-consistent review of the standard thermodynamics of weakly-coupled open quantum systems, that emphasizes the approximations taken, which no longer hold for the non-Markovian case. The thermodynamics formulation for the refined coupling limit is presented in Sec. 4, and constitutes the main result of this work. We complete our study by applying these ideas to the case of a two-level system in contact with a thermal bath in Sec. 5. Finally, some discussion and possible future directions of this approach are outlined in the Conclusions section.

\section{Weak Coupling and Refined Weak Coupling Limit}
Let $\mathscr{S}$ be an open system in contact with some environment $\mathscr{E}$. The total Hamiltonian for system and environment is given by an expression of the form $H=H_{\mathscr{S}}+H_{\mathscr{E}}+V_{\mathscr{SE}}$, where $H_{\mathscr{S}}$, $H_{\mathscr{E}}$, and $V_{\mathscr{SE}}$ denote the system, environment, and interaction Hamiltonian, respectively. In the interaction picture, the von Neumann equation reads:
\begin{equation}\label{VNeq}
\frac{d\tilde{\rho}_{\mathscr{SE}}(t)}{dt}=-\ii [\tilde{V}_{\mathscr{SE}}(t),\tilde{\rho}_{\mathscr{SE}}(t)],
\end{equation}
where, for any operator $A$, $\tilde{A}(t):=\exp[\ii (H_{\mathscr{S}}+H_{\mathscr{E}})t]A\exp[-\ii (H_{\mathscr{S}}+H_{\mathscr{E}})t]$, and we have taken units of $\hbar=1$. Unless otherwise stated, we shall consider units of $\hbar=k_{\rm B}=1$ in the following. In addition, we shall take the initial time $t=0$ to be the time where system and environment start interacting, $\rho_{\mathscr{SE}}(0)=\rho_{\mathscr{S}}(0)\otimes\rho_{\mathscr{E}}(0)$, so that the system dynamics can be described in terms of a dynamical map $\Lambda_t$, i.e., a linear, trace preserving, and CP map. 

\subsection{Weak Coupling Limit}
In order to obtain $\Lambda_t$ from \eqref{VNeq}, some assumptions are usually required, the weak coupling case being the most common one. In the standard approach to this situation, the strength of the interaction is gauged by some coupling constant $\alpha$, so that $V_{\mathscr{SE}}$ is substituted by $\alpha V_{\mathscr{SE}}$ in \eqref{VNeq}, and the limit $\alpha\to 0$ is taken on a rescaled time $\tau\to\alpha^2 t$. This leads to a GKLS equation \cite{Davies,AlickiBook,Libro}:
\begin{equation}
\frac{d\tilde{\rho}_\SS(\tau)}{d\tau}:=\tilde{\mathcal{L}}[\tilde{\rho}_\SS(\tau)]=\ii [H_{\rm LS},\tilde{\rho}_\SS(\tau)]+\sum_{\omega}\sum_{k,l}\gamma_{kl}(\omega)\left[A_l(\omega)\tilde{\rho}_\SS(\tau)A_{k}^\dagger -\tfrac12 \big\{A_k^\dagger(\omega) A_l(\omega),\tilde{\rho}_{\SS}(\tau)\big\}\right].
\end{equation}
Here, $\tilde{\mathcal{L}}$ is the Liouville operator (in the interaction picture), and $A_k(\omega)$ are eigenoperators of the system Hamiltonian, $[H_{\mathscr{S}},A_k(\omega)]=-\ii\omega A_k(\omega)$, such that the interaction can always be written in the form of $V_{\mathscr{SE}}=\sum_k\sum_{\omega} A_k(\omega)\otimes B_k$ with $B_k$ Hermitian bath operators (see, e.g., \cite{Libro}). Moreover, $H_{\rm LS}$ is a Hamiltonian Lamb-shift-type correction $[H_{\mathscr{S}},H_{\rm LS}]=0$, and $\gamma_{kl}(\omega)$ are the elements of a positive-semidefinite matrix (see the details in \cite{BrPr02,AlickiBook,Libro}). In addition, $\rho_{\mathscr{E}}(0)$ is assumed to be some Gaussian state of an environment with infinitely many degrees of freedom and $\Tr_{\mathscr{E}}[V_{\mathscr{SE}}\rho_{\mathscr{E}}(0)]=0$. This last condition can always be achieved by a proper redefinition of system and interaction Hamiltonians \cite{Libro}. Under some relatively mild conditions on the environmental correlation functions, Davies rigorously shows \cite{Davies}: 
\begin{equation}\label{DaviesTheorem}
\lim_{\alpha\to 0} \sup_{0\leq \alpha^2 t \leq t_0} \|\tilde{\Lambda}_t(\rho_{\SS})-\ee^{\alpha^2\tilde{\mathcal{L}}t}(\rho_{\SS})\|=0, \quad t_0<\infty,
\end{equation}
for any initial system state $\rho_{\SS}$. Here, $\tilde{\Lambda}_t$ denotes the exact dynamical map (in the interaction picture). Recently, precise bounds to this convergence have been formulated \cite{Merkli}. The Davies' result allows for the approximation
\begin{equation}
\Lambda_t\simeq \ee^{(-\ii\mathcal{H}+\alpha^2\mathcal{L})t}:= \ee^{\mathcal{L}_{\rm D}t}
\end{equation}
for $t\lesssim\alpha^{-2}$. Here, $\mathcal{L}_{\rm D}$ is the so-called Davies generator, $\mathcal{H}:= [H_{\mathscr{S}},\cdot]$, working again in the Schr\"odinger picture, and the relation $[\mathcal{H},\mathcal{L}]=0$ is fulfilled. This is the celebrated weak coupling limit. 

Furthermore, this technique can be extended for time-dependent Hamiltonians $H_{\SS}(t)$. If $\tau_H$ is the typical time for the variation of $H_{\SS}(t)$, in the simplest case, this change is considered to be slow in comparison to the evolution of the open system due to the coupling to the environment, i.e., $\tau_H \gtrsim \alpha^{-2}$ \cite{DaviesSpohn}. Then, we can ``adiabatically'' deform the generator $\mathcal{L}_{\rm D}\to \mathcal{L}_{\rm D}(t)$, so that $\mathcal{L}_{\rm D}(t)$ is the Davies generator calculated instantaneously for Hamiltonian $H_{\SS}(t)$, and the approximation
\begin{equation}\label{DaviesAd}
\Lambda_t\simeq \mathcal{T}\ee^{\int_{0}^{t}\mathcal{L}_{\rm D}(s)ds}
\end{equation}
is satisfied, where $\mathcal{T}$ is the time-ordering operator. We should note that the term ``adiabatic'' is used throughout the text with its standard meaning in quantum mechanics referring to slow transformations. We do not intend to denote lack of heat exchange as in classical thermodynamics.

\subsection{Refined Weak Coupling Limit}

Despite enjoying very remarkable properties, the Davies' weak coupling technique has a problem for very short times $t\ll \alpha^{-2}$, as in this case, $\tilde{\Lambda}_t\simeq \exp(\alpha^2\tilde{\mathcal{L}}t)\simeq \mathds{1}$, and no open system dynamics can be resolved. In order to see this short time scale, a \emph{refined} weak coupling must be formulated \cite{AlickiRefined,SchallerBrandes,Benatti1,Refined,Libro}. Nonetheless, one has to be careful to prevent violations of complete positivity \cite{noCP1,noCP2,noCP3}. To this end, we can solve \eqref{VNeq} as a time-ordered exponential:
\begin{align}\label{Tord}
\tilde{\rho}_{\mathscr{SE}}(t)=\mathcal{T}\ee^{\int_0^t \tilde{\mathcal{V}}(s)ds}[\rho_{\mathscr{S}}(0)\otimes\rho_{\mathscr{E}}(0)],
\end{align}
with $\tilde{\mathcal{V}}(t):=-\ii [\tilde{V}(t),\cdot]$, and look for a solution of the reduced dynamics in the form of
\begin{align}
\tilde{\rho}_\SS(t)=\tilde{\Lambda}_t[\rho_\SS(0)]=\ee^{\mathcal{Z}(t)}[\rho_\SS(0)],
\end{align}
where the exponent $\mathcal{Z}(t)$ is formally defined as the logarithm of $\tilde{\Lambda}_t$. Since $\tilde{\Lambda}_t$ can be expressed as a power expansion in $\alpha$, we can assume a similar expansion for the exponent:
\begin{equation}
\mathcal{Z}(t)=\sum_{k=1}^\infty \alpha^k \mathcal{Z}_k(t).
\end{equation}
The terms $\mathcal{Z}_k(t)$ can be computed expanding the exponential and by comparison with the terms of the same exponent of $\alpha$ in the expansion of $\eqref{Tord}$ after taking the partial trace on the environment \cite{AlickiRefined,SchallerBrandes,Benatti1,Refined}. The first order vanishes because $\Tr_{\mathscr{E}}[V_{\mathscr{SE}}\rho_{\mathscr{E}}(0)]=0$, so that the first nontrivial contribution is 
\begin{equation}
\mathcal{Z}_2(t)[\rho_\SS(0)]=-\frac{1}{2}\mathcal{T}\int_0^tdt_1\int_0^tdt_2\Tr_{\mathscr{E}}\left[\tilde{V}_{\mathscr{SE}}(t_1),\left[\tilde{V}_{\mathscr{SE}}(t_2),\rho_\SS(0)\otimes\rho_{\mathscr{E}}(0)\right]\right].
\end{equation}
Applying $\mathcal{T}$ under the integral signs and reordering terms, we obtain
\begin{align}\label{DCG2}
\mathcal{Z}_2(t)[\rho_\SS(0)]=-\ii[\Lambda(t),\rho_\SS(0)]+\Tr_E\big[\Upsilon(t)\rho_\SS(0)\otimes\rho_{\mathscr{E}}(0) \Upsilon(t)-\tfrac{1}{2}\big\{\Upsilon^2(t),\rho_\SS(0)\otimes\rho_{\mathscr{E}}(0)\big\}\big],
\end{align}
with Hermitian operators $\Lambda(t):=\tfrac{1}{2\ii}\int_0^tdt_1\int_0^tdt_2\sgn(t_1-t_2)\Tr_E[\tilde{V}_\SE(t_1)\tilde{V}_\SE(t_2)\rho_{\mathscr{E}}(0)]$ and $\Upsilon(t):=\int_0^t\tilde{V}_\SE(s)ds$. In terms of eigenoperators of $H_{\SS}$, this is written as \cite{AlickiRefined,Refined}
\begin{align}\label{Z}
  \mathcal{Z}_2(t)[\rho_\SS(0)]=&-\ii[\Lambda(t),\rho_\SS(0)] \nonumber \\
  &+\sum_{\omega,\omega'}\sum_{k,l} \Gamma_{kl}(\omega,\omega',t)\big[A_l(\omega')\rho_\SS(0) A_k^\dagger(\omega)-\tfrac{1}{2}\{A_k^\dagger(\omega) A_l(\omega'),\rho_\SS(0) \}\big], 
\end{align}
where
\begin{align}\label{GAMMAkl}
\Gamma_{kl}(\omega,\omega',t):=\int_{0}^{t}dt_1\int_{0}^{t}dt_2 \ee^{\ii(\omega t_1-\omega' t_2)}\Tr[\tilde{B}_k(t_1-t_2)B_l\rho_\EE(0)],
\end{align}
and
\begin{equation}
\Lambda(t):=\sum_{\omega,\omega'}\sum_{k,l}\Xi_{kl}(\omega,\omega',t)A_k^\dagger(\omega) A_l(\omega'),
\end{equation} 
with
\begin{equation}\label{XIkl}
\Xi_{kl}(\omega,\omega',t):=\frac{1}{2\ii}\int_{0}^{t}dt_1\int_{0}^{t}dt_2\sgn(t_1-t_2)\ee^{\ii(\omega t_1-\omega' t_2)}\Tr[\tilde{B}_k(t_1-t_2)B_l\rho_\EE(0)].
\end{equation}

From \eqref{DCG2}, we infer that $\mathcal{Z}_2(t)$ has the GKLS form \cite{GKLS1,GKLS2}, so it turns out that the coefficients $\Gamma_{kl}(\omega,\omega',t)$ form a positive-semidefinite matrix. This property seems not ensured beyond the second order \cite{Schaller2}.

Thus, the refined weak coupling limit is performed by neglecting higher orders on $\alpha$ in the exponent, $\mathcal{Z}(t)\simeq \mathcal{Z}_2(t)$:
\begin{equation} \label{refi1}
\Lambda_t=\ee^{-\ii\mathcal{H}t}\ee^{\mathcal{Z}(t)}\simeq \ee^{-\ii\mathcal{H}t}\ee^{\alpha^2\mathcal{Z}_2(t)}.
\end{equation}
By construction, this is a completely positive map [$\mathcal{Z}_2(t)$ has the GKLS form for all $t$] which approaches the exact dynamics in the short time limit $t\ll \alpha^{-2}$ (note that, in general, $[\mathcal{Z}_2(t),\mathcal{H}]\neq0$). In the large time scale, $t\sim \alpha^{-2}$, it can be proven \cite{AlickiRefined,SchallerBrandes,Refined} that $\alpha^2\mathcal{Z}_2(t)\simeq \tilde{\mathcal{L}}_{\rm D} t$, and the refined weak coupling dynamics $\Lambda_t^{\rm R}$ approaches the weak coupling limit Davies' semigroup,
\begin{equation}\label{REFapproxDavies}
\Lambda_t^{\rm R}:=\ee^{-\ii\mathcal{H}t}\ee^{\alpha^2\mathcal{Z}_2(t)}\simeq \ee^{-\ii\mathcal{H}t}\ee^{\alpha^2\tilde{\mathcal{L}}_{\rm D}t}=\ee^{\mathcal{L}_{\rm D}t}, \quad t\sim  \alpha^{-2}.
\end{equation}
Hence, thanks to the Davies' result \eqref{DaviesTheorem}, the dynamical map $\Lambda_t^{\rm R}$ provides a nontrivial and consistent approximation for the weak coupling dynamics for the same time scale as the Davies' semigroup, $0\leq t \lesssim \alpha^{-2}$. However, $\Lambda_t^{\rm R}$ resolves the exact dynamics in the small time scale. One may notice that under the assumption that the width of the reservoir correlation functions is negligible in comparison with the time scale of the open system, $\exp(\mathcal{L}_{\rm D}t)$ and $\Lambda_t^{\rm R}$ coincide as the integrals in Eqs. \eqref{GAMMAkl} and \eqref{XIkl} can be effectively extended to infinity for all times.

We can calculate the generator $\mathcal{L}_{\rm R}(t)$ of $\Lambda_t^{\rm R}$, i.e., $\frac{d}{dt}\Lambda_t^{\rm R}=\mathcal{L}_{\rm R}(t)\Lambda_t^{\rm R}$, by applying inverses and derivatives,
\begin{equation}\label{Lr1}
\mathcal{L}_{\rm R}(t)=\left(\frac{d}{dt}\Lambda_t^{\rm R}\right)(\Lambda_t^{\rm R})^{-1}=-\ii \mathcal{H}+\ee^{-\ii\mathcal{H}t}\left[\frac{d}{dt}\ee^{\alpha^2\mathcal{Z}_2(t)}\right]\ee^{-\alpha^2\mathcal{Z}_2(t)}\ee^{\ii\mathcal{H}t}=-\ii \mathcal{H}+\ee^{-\ii\mathcal{H}t}\tilde{\mathcal{L}}_{\rm R}(t)\ee^{\ii\mathcal{H}t},
\end{equation}
where $\tilde{\mathcal{L}}_{\rm R}(t)$ is the generator in the interaction picture that can be written in terms of the time derivative of $\mathcal{Z}_2(t)$ with the help of the Snider-Wilcox formula \cite{SWformula},
\begin{equation}\label{Snider-Wilcox}
\frac{d}{dt}\ee^{A(t)}=\int_0^1 \ee^{s A(t)}\left[\frac{d}{dt}A(t)\right]\ee^{(1-s) A(t)}ds,
\end{equation}
so that
\begin{equation}
\tilde{\mathcal{L}}_{\rm R}(t)=\left[\frac{d}{dt}\ee^{\alpha^2\mathcal{Z}_2(t)}\right]\ee^{-\alpha^2\mathcal{Z}_2(t)}=\alpha^2\int_0^1 \ee^{s \alpha^2\mathcal{Z}_2(t)}\dot{\mathcal{Z}}_2(t) \ee^{-s \alpha^2\mathcal{Z}_2(t)}ds.
\end{equation}
Here, we have used the ``overdot'' notation for the time derivative. Since $\alpha^2\mathcal{Z}_2(t)\simeq \tilde{\mathcal{L}}_{\rm D} t$, from Eq. \eqref{Lr1}, it is straightforward to check that for large $t$, $\mathcal{L}_{\rm R}(t)\simeq \mathcal{L}_{\rm D}$.

Although we can think of the refined weak coupling as a small correction to the Markovian weak coupling evolution for short times, in general, the dynamical map $\Lambda_t^{\rm R}$ turns out to be highly non-Markovian \cite{noMarkov1,noMarkov2,noMarkov3}, breaking conditions such as P-divisibility \cite{Refined}.

\subsection{Refined Weak Coupling Limit under Slowly-Varying Time-Dependent Hamiltonians}
As far as we know, the refined weak coupling method for a time-dependent system Hamiltonian $H_\SS(t)$ has not been properly studied in the literature. A priori, the simplest situation should be again the case of slowly-varying $H_\SS(t)$, i.e., $\tau_H \gtrsim \alpha^{-2}$, but
the problem becomes now more involved than in the standard weak coupling. We would expect an ``adiabatic deformation'' of the refined weak coupling generator such that it approaches the aforementioned adiabatically-deformed Davies generator $\mathcal{L}_{\rm D}(t)$ for large times. However, it is not straightforward to obtain this result keeping CP. 

In order to do so, notice that, on the one hand, if $\mathcal{Z}_2(t,s)$ is the refined weak coupling exponent calculated for the Hamiltonian $H_{\SS}(s)$, for $s$ fixed, we expect $\alpha^2\mathcal{Z}_2(t,s)\simeq \tilde{\mathcal{L}}_{\rm D}(s)t$ for large $t$ (any $s$). On the other hand, we may observe that, in the weak coupling limit, the substitution
\begin{equation}
\ee^{\mathcal{L}_{\rm D}t}\to \mathcal{T}\ee^{\int_0^t\mathcal{L}_{\rm D}(s)ds}
\end{equation} 
when passing from $H_{\SS}$ to $H_{\SS}(t)$ in the adiabatic approximation can be seen as a kind of ``averaging'' approximation. Namely, we construct $\mathcal{L}_{\rm D}(s)$ for $H_{\SS}(s)$, for $s$ fixed, and approximate the evolution as 
\begin{equation}
\mathcal{T}\ee^{\overline{\mathcal{L}_{\rm D}(s) t}}= \mathcal{T}\ee^{\int_0^t \mathcal{L}_{\rm D}(s) ds},
\end{equation}
where the temporal average of some function of $s$, $f(s)$, has been introduced $\overline{f(s)}:=\frac{1}{t}\int_0^t f(s)ds$. Thus, the evolution from $0$ to $t$ is approximately given, modulo time-ordering, by the exponential of the ``time-averaged'' value of the exponent $\mathcal{L}_{\rm D}(s) t$ from $0$ to $t$.

Similarly, in the refined weak coupling limit, we can perform an adiabatic approximation for slowly-varying  $H_{\SS}(t)$ by 
\begin{equation}\label{LambdaRA1}
\tilde{\Lambda}_t^{\rm RA}:= \mathcal{T}\ee^{\overline{\alpha^2\mathcal{Z}_2(t,s)}} =\mathcal{T} \ee^{\frac{\alpha^2}{t}\int_0^{t}\mathcal{Z}_2(t,s) ds},
\end{equation}
which in the Schr\"odinger picture yields
\begin{equation}\label{LambdaRA2}
\Lambda_t^{\rm RA}=\mathcal{U}_t\tilde{\Lambda}_t^{\rm RA}, \quad\text{with} \quad \mathcal{U}_t:=\mathcal{T}\ee^{-\ii\int_0^t\mathcal{H}(s) ds}.
\end{equation}
As required, $\Lambda_t^{\rm RA}$ approaches \eqref{DaviesAd} for large $t$ and reduces to $\Lambda_t^{\rm R}$ for time-independent system Hamiltonians. Furthermore, $\Lambda_t^{\rm RA}$ is, by construction, a valid CP map because $\mathcal{Z}_2(t,s)$ has the GKLS form for any $t$ and $s$ (and $t>0$).

\section{Standard Thermodynamics of Open Quantum Systems}
We shall assume from now on the environment is a thermal reservoir or bath. Namely, $\mathscr{E}$ is a system with infinitely many degrees of freedoms, which remains initially in the canonical Gibbs state $\rho_{\mathscr{E}}=\rho_\mathscr{E}^\beta=Z_{\EE}^{-1}\exp(-\beta H_{\mathscr{E}})$. If the system Hamiltonian $H_\SS$ is time independent, the system Gibbs state $\rho_\mathscr{S}^\beta=Z_{\SS}^{-1}\exp(-\beta H_{\mathscr{S}})$ is a steady state of the Davies semigroup \cite{AlickiBook,BrPr02,Libro,Davies},
\begin{equation}
\mathcal{L}_{\rm D}\big(\rho_\mathscr{S}^\beta \big)=0\Rightarrow \ee^{\mathcal{L}_{\rm D}t}\big(\rho_\mathscr{S}^\beta \big)=\rho_\mathscr{S}^\beta.
\end{equation}
In fact, provided that system-reservoir coupling $V_{\SE}$ only allows $\rho_\mathscr{S}^\beta$ as a steady state, the evolution of any system eventually approaches this unique steady state \cite{AlickiBook,Libro,BrPr02,Spohn2,SpohnLebowitz,Wolf},
\begin{equation}
\lim_{t\to\infty }\ee^{\mathcal{L}_{\rm D}t}\big[\rho_\mathscr{S}(0)\big]=\rho_\mathscr{S}^\beta.
\end{equation} 
In the case of a slowly-varying time-dependent system Hamiltonian $H_{\SS}(t)$, the adiabatically-deformed Davies $\mathcal{L}_{\rm D}(t)$ generator fulfills
\begin{equation}\label{SSCanonicalt}
\mathcal{L}_{\rm D}(t)\Big[\rho_{\SS(t)}^\beta \Big]=0, \quad \text{with}\quad  \rho_{\SS(t)}^\beta=Z_{\SS (t)}^{-1}\ee^{-\beta H_{\mathscr{S}}(t)}.
\end{equation}
Provided that $\rho_{\SS(t)}^\beta$ is the only state satisfying this, and $H_\SS(t)$ remains almost constant during the relaxation time --very slow variation limit--, for large enough $t$, any initial state approaches $\rho_{\SS(t)}^\beta$,
\begin{equation}\label{Ergodict}
\mathcal{T}\ee^{\int_0^t \mathcal{L}_{\rm D}(s)ds}\big[\rho_\mathscr{S}(0)\big]\simeq \rho_{\SS (t)}^\beta.
\end{equation}

According to the first law of thermodynamics, the change of the internal energy of $\mathscr{S}$ can be divided into work and heat. The former is usually identified with a controllable and measurable change, for instance by means of variations in the parameters of $H_{\mathscr{S}}$ (which must be then time-dependent), whereas the latter is a change generally out of our capability to control and observe, and it is often identified with reservoir properties. However, the specific definition of work and heat is in general difficult because the identification of internal energy  for a given system-reservoir Hamiltonian $H(t)=H_{\SS}(t)+H_{\EE}+\alpha V_{\SE}$ is also difficult. It is natural to include $H_{\SS}(t)$ inside the system internal energy, but it is not clear which part of the interaction term $\alpha V_{SE}$ should be considered as ``internal'' to the system. In general, we have
\begin{equation}\label{<H>}
\langle H(t)\rangle=\langle H_\SS(t)\rangle+\langle H_\EE\rangle+\alpha \langle V_{SE}\rangle=\Tr[\rho_\SS(t)H_\SS(t)]+\Tr[\rho_\EE(t)H_\EE]+\alpha\Tr[\rho_\SE(t)V_\SE].
\end{equation} 
Nevertheless, in the weak coupling limit $\alpha^2\to0$, 
\begin{equation}\label{weakEnergysplit}
\langle H(t)\rangle\simeq\Tr[H_{\mathscr{S}}(t)\rho_{\mathscr{S}}(t)]+\Tr[H_\EE\rho_\EE(t)],
\end{equation}
hence the internal energy can be defined via
\begin{align}
E(t):=\Tr[\rho_{\mathscr{S}}(t)H_{\mathscr{S}}(t)]. \label{EintWeak}
\end{align}
Actually, under the condition $\Tr_\EE(V_\SE\rho_\EE^\beta)=0$, it can be proven that the neglected term in \eqref{weakEnergysplit} is order $\alpha^2$, instead of order $\alpha$ (see Appendix \ref{A}). 

\subsection{The First Law}

Taking the time derivative in the internal energy \eqref{EintWeak},
\begin{equation}
\dot{E}(t)=\Tr[\dot{\rho}_{\mathscr{S}}(t)H_{\mathscr{S}}(t)]+\Tr[\rho_{\mathscr{S}}(t) \dot{H}_{\mathscr{S}}(t)],
\end{equation}
and heat and work become defined via its time derivative:
\begin{align}
\dot{Q}(t):=\Tr[\dot{\rho}_{\mathscr{S}}(t)H_{\mathscr{S}}(t)],\label{Qweak}\\
\dot{W}(t):=\Tr[\rho_{\mathscr{S}}(t) \dot{H}_{\mathscr{S}}(t)],\label{Wweak}
\end{align}
so that
\begin{equation}\label{1law}
\dot{E}(t)=\dot{Q}(t)+\dot{W}(t),
\end{equation}
with the heat flow $\dot{Q}(t)$ into the system and applied power $\dot{W}(t)$ at time $t$ positive if they increase the system energy. Consequently, the integrated form of the first law reads
\begin{equation}\label{1lawI}
\Delta E(t)=Q(t)+W(t),
\end{equation}
with $\Delta E(t):=E(t)-E(0)$, and
\begin{align}
Q(t)=\int_0^{t}\Tr[\dot{\rho}_{\mathscr{S}}(r)H_{\mathscr{S}}(r)]dr,\label{QweakI}\\
W(t)=\int_0^{t}\Tr[\rho_{\mathscr{S}}(r) \dot{H}_{\mathscr{S}}(r)]dr.\label{WweakI}
\end{align}

\subsection{The Second Law}

In order to derive the second law, we define the thermodynamic entropy of the system by
\begin{equation}
S(t):=k_BS[\rho_{\SS}(t)],
\end{equation}
where $S(\rho)=-\Tr(\rho\log\rho)$ is the von Neumann entropy, so that
\begin{equation}
S(t)=-\Tr[\rho_{\SS}(t)\log \rho_{\SS}(t)],
\end{equation}
in units of $k_B=1$. The quantum relative entropy between two states $\rho_1$ and $\rho_2$ is defined by
\begin{equation}
S(\rho_1\|\rho_2):=\Tr(\rho_1\log\rho_1)-\Tr(\rho_1\log\rho_2).
\end{equation}
It can be proven \cite{Monotonic} that $S(\rho_1\|\rho_2)$ is monotonic under any CP and trace preserving map $\Phi$, 
\begin{equation}\label{monotonicity}
S[\Phi(\rho_1)\|\Phi(\rho_2)]\leq S(\rho_1\|\rho_2).
\end{equation}
Recently, the proof has been extended to any positive and trace preserving map $\Phi$ \cite{PMonotonic}. Suppose $\mathcal{L}(t)$ is any time-dependent GKLS generator with a steady state $\rho_{\rm ss}(t)$, $\mathcal{L}(t)[\rho_{\rm ss}(t)]=0$. Considering the CP map $\exp[\mathcal{L}(t)r]$ with $r\geq0$, we have
\begin{equation}
S[\ee^{\mathcal{L}(t) r}(\rho)\|\rho_{\rm ss}(t)]=S\{\ee^{\mathcal{L}(t) r}(\rho)\|\ee^{\mathcal{L}(t) r}[\rho_{\rm ss}(t)]\}\leq S[\rho\|\rho_{\rm ss}(t)], \quad r>0,
\end{equation}
which implies
\begin{equation}
\frac{d}{dr}S[\ee^{\mathcal{L}_t r}(\rho)\|\rho_{\rm ss}(t)]\leq 0,
\end{equation}
and, particularly for $r=0$, the inequality \cite{SpohnEntropy}
\begin{equation}\label{Spohn}
\Tr\big([\mathcal{L}_t(\rho)]\big\{\log(\rho)-\log[\rho_{\rm ss}(t)]\big\}\big)\leq0.
\end{equation}
Here, for the time derivative of the thermodynamic entropy, one has the general result \cite{Wilde}:
\begin{equation}\label{Sdiff}
\dot{S}(t)=-\frac{d}{dt}\Tr[\rho_{\SS}(t)\log \rho_{\SS}(t)]=-\Tr\left[\dot\rho_{\SS}(t)\log \rho_{\SS}(t)\right].
\end{equation}
The bound \eqref{Spohn} is sometimes often called Spohn's inequality \cite{QthermoKosloff,Qthermo2,AlickiTutorialPeriodic,StrasbergPRX}. In the (adiabatic) weak coupling limit \eqref{DaviesAd}, Eq. \eqref{Sdiff} becomes     
\begin{equation}\label{Sprima}
\dot{S}(t)=-\Tr\left\{\left[\mathcal{L}_{\rm D}(t)\rho_{\SS}(t)\right]\log \rho_{\SS}(t)]\right\}.
\end{equation}
Since the canonical Gibbs state $\rho_{\SS(t)}^\beta$ is a steady state of $\mathcal{L}_{\rm D}(t)$, Eq. \eqref{SSCanonicalt}, the inequality \eqref{Spohn} combined with \eqref{Sprima} and \eqref{Qweak} leads to
\begin{equation}\label{2lawweak}
\dot{S}(t)-\beta \dot{Q}(t)\geq0.
\end{equation}
This is the differential form of the second law of nonequilibrium thermodynamics in the weak coupling limit. It implies the integrated form:
\begin{align}\label{2law}
\Delta{S}(t)-\beta {Q}(t)\geq0,
\end{align}
with $\Delta S(t)=S(t)-S(0)$. Namely, the entropy production [here also its rate \eqref{2lawweak}] due to the interaction with the reservoir cannot be negative for any final time $t$.

\subsection{Difficulties beyond the Weak Coupling Limit}

When the interaction term $\alpha V_{\SE}$ cannot be neglected, we cannot expect inequality \eqref{2lawweak} to be satisfied. In general, the open system evolution becomes non-Markovian \cite{noMarkov1,noMarkov2,noMarkov3}, and the generator of such a dynamics, $\mathcal{L}(t)$, provided it is well defined, does not have the GKLS form. Spohn's inequality \eqref{Spohn} cannot be applied. 

The problems go a step back because, as aforementioned, it is not clear which part of the total energy $\langle H (t)\rangle=\langle H_{\SS}(t)\rangle+\langle H_{\EE}(t)\rangle+\alpha \langle V_{\SE}(t)\rangle$ must be considered as internal energy of the system. A possible ``extreme-type'' of splitting is $\dot{E}(t)=\Tr[(H_{\SS}+\alpha V_{\SE})\dot{\rho}_{\SE}(t)]$, and so $\dot{Q}(t)=\Tr[H_{\EE}\dot{\rho}_{\SE}(t)]$; then, it can be proven that the inequality \eqref{2law} is always satisfied \cite{StrasbergPRX,Esposito,Tanimura}. However, it is satisfied for a reservoir of any size, and this might generate some criticism reading its strict equivalence to the phenomenological second law of nonequilibrium thermodynamics. 

Another strategy for non-negligible $\alpha$ in the case of time-independent $H_{\SS}$ is based on the fact that 
\begin{equation}\label{GibbsSE}
\rho_{\SE}^\beta=\frac{\exp(-\beta H)}{Z_{\SE}}=\frac{\exp[-\beta (H_{\SS}+H_{\EE}+\alpha V_{\SE})]}{Z_{\SE}}
\end{equation}
is the global equilibrium state of system and reservoir. Therefore, the reduced state
\begin{equation}
\rho_{\SS}^{\rm ss}=\Tr_{\EE}\big(\rho_{\SE}^\beta\big)
\end{equation}
is a fixed point of the system dynamics. Formally, one can write this reduced state as a Gibbs state 
\begin{equation}\label{H*}
\rho_{\SS}^{\rm ss}=\frac{\exp(-\beta H_{\SS}^*)}{Z_{\SS}^*}, \quad \text{with \ } H_{\SS}^*:=-\beta^{-1}\log \Tr_{\EE}\big(\rho_{\SE}^\beta\big)
\end{equation}
a Hamiltonian of ``mean'' force \cite{Gelin,Seifert,Hu}. This suggests a possible choice of internal energy as 
\begin{equation}
\tilde{E}^*(t):=\Tr[H_{\SS}^*\rho_{\SS}(t)],
\end{equation}
which approaches the weak coupling internal energy as $\alpha\to0$, Eq. \eqref{EintWeak}. Then, in absence of work, $\tilde{Q}=\Tr[H_{\SS}^*\rho_{\SS}(t)]$. However, this choice does not allow for the use of the monotonicity of the relative entropy \eqref{monotonicity} to obtain the second law. If the system is initially in the state $\rho_{\SS}^{\rm ss}$, it remains invariant provided that the initial system-reservoir state is in the total canonical ensemble \eqref{GibbsSE}. Since this is not a product state, the reduced dynamics would not be given by a general (CP) dynamical map.

A slightly different choice, originally due to Seifert \cite{Seifert}, defines internal energy at equilibrium by using a classical thermodynamic relation:
\begin{equation}\label{E*}
E^*:=-\frac{\partial}{\partial \beta}\log Z_{\SS}^*=\Tr\big[\rho_{\SS}^{\rm ss}\left(H_{\SS}^*+\beta\partial_{\beta}H_{\SS}^*\right)\big],
\end{equation}
which suggests the out of equilibrium definition \cite{StrasbergPRE}:
\begin{equation}\label{E*(t)}
E^*(t):=\Tr\big\{\rho_{\SS}(t)\left[H_{\SS}^*(t)+\beta\partial_{\beta}H_{\SS}^*(t)\right]\big\},
\end{equation}
with $H_{\SS}^*(t)$ defined as $H_{\SS}^*$ in Eq. \eqref{H*} for the canonical Gibbs state with a time-dependent $H_{\SS}(t)$. In such a case, the classical definition of free energy $F:=-\beta^{-1}\log Z_{\SS}$ jointly with the relation $F=E-T S$, motivates the following redefinition of free energy and thermodynamic entropy:
\begin{align}
&F^*(t):=\Tr\big\{\rho_{\SS}(t)\big[H_{\SS}^*+\beta^{-1}\log\rho_{\SS}(t)\big]\big\},\\
&S^*(t):=\Tr\big\{\rho_{\SS}(t)\big[-\log\rho_{\SS}(t)+\beta^2\partial_{\beta}H_{\SS}^*\big]\big\},
\end{align}
so that $F^*=E^*-TS^*$. It is possible to obtain an equation with the form of \eqref{2law} for the dynamics with initial system-reservoir states to be either the total canonical ensemble \eqref{GibbsSE} or belonging to a class of zero discord states \cite{StrasbergPRE}.

Other approaches beyond the weak coupling framework that are based on e.g., reservoir full counting statistics or coordinate mappings can be found in Chaps. 11 and 15, and 22-25 of \cite{Qthermo2} and the references therein, respectively.

\section{Thermodynamics in the Refined Weak Coupling Limit}
The analysis of thermodynamic properties in the refined weak coupling limit is challenging because of several reasons. Since the $\Lambda_t^{\rm R}$ is in general non-Markovian, the breaking of P-divisibility precludes the use of the Spohn's inequality \eqref{Spohn} to formulate the second law \cite{Pati}. For non-Markovian dynamics, where there is a back-and-forth of information between system and environment, it is reasonable to consider that global system-reservoir information measures must be included in the second law \cite{Esposito,Alipour2,StrasbergPRX}. However, we will take a different approach here.

Moreover, in the refined weak coupling limit, we cannot neglect the interaction term $\alpha V_{\SE}$ in the total energy \eqref{<H>} for short times. This can be seen with the following argument. If the initial system-reservoir state is taken to be
\begin{equation}\label{RWCinitial}
\rho_{\SE}(0)=\rho_{\SS}^{\beta}\otimes \rho_{\EE}^{\beta},
\end{equation}
under the hypothesis that the $\alpha V_{\SE}$ is negligible, we have for the total Gibbs state \eqref{GibbsSE} 
\begin{equation}\label{canonicalfactorize}
\rho_{\SE}^\beta=\frac{\exp(-\beta H)}{Z_{\SE}}\simeq \frac{\exp(-\beta H_{\SS})}{Z_{\SS}}\otimes \frac{\exp(-\beta H_{\EE})}{Z_{\EE}}=\rho_{\SE}(0).
\end{equation}
Since \eqref{GibbsSE} is a stationary state of the total dynamics, Eq. \eqref{canonicalfactorize} would imply $\rho^{\beta}_{\SS}$ is a steady state of the reduced system dynamics in the refined weak coupling limit. However, $\rho^{\beta}_{\SS}$ is not a steady state in the refined weak coupling for finite $t$,
\begin{equation}\label{RWCinfinityNEQ}
\Lambda_t^{\rm R}\rho^{\beta}_{\SS}\neq \rho^{\beta}_{\SS},
\end{equation}
but
\begin{equation}\label{RWCinfinity}
\lim_{t\to\infty}\Lambda_t^{\rm R}\rho^{\beta}_{\SS}= \rho^{\beta}_{\SS}.
\end{equation}
Hence, the identifications \eqref{EintWeak} and \eqref{Qweak} for internal energy and heat in absence of work, respectively, cannot be assumed to be true in the refined weak coupling limit for finite times. 

\subsection{Time-Independent System Hamiltonian $H_{\SS}$}
Let us consider first the case of a time-independent system Hamiltonian $H_{\SS}$, so that no work is applied to/performed by the system. We could adopt the definition $E^*(t)$ \eqref{E*(t)} for internal energy. This has the drawback that, since \eqref{EintWeak} does not coincide with \eqref{E*} unless the weak interaction term $\alpha V_{\SE}$ is neglected,  the equilibrium internal energy in the refined weak coupling would not coincide with the weak coupling one \eqref{EintWeak} for large times, despite the fulfillment of \eqref{RWCinfinity}. In addition, the refined weak coupling limit assumes an initial system-reservoir product state, which implies $\langle H(0) \rangle=\langle H_{\SS}(0)\rangle+\langle H_{\EE}(0)\rangle$, as assumed $\Tr_{\EE}[V_{\SE}\rho_{\EE}(0)]=0$. Then, the internal energy at $t=0$ must be $\langle H_{\SS}(0)\rangle$, and this is not obtained with the choice \eqref{E*}. Therefore, by construction, the internal energy in the refined weak coupling regime should be defined such that it fits the weak coupling internal energy $E(t)=\Tr[\rho_{\SS}(t) H_\SS (t)]$ at $t=0$ and for asymptotic times, but differs from $E(t)$ at finite times.

Such a definition is nevertheless possible by following a similar argument to the Hamiltonian of a mean force \eqref{H*}. That is, we can write the time-evolution of the system Gibbs state $\rho_{\SS}^\beta=Z_{\SS}^{-1}\exp(-\beta H_{\SS})$ in a canonical form:
\begin{equation}
\Lambda_{t}^{\rm R}\left(\rho_{\SS}^\beta\right)=\frac{\ee^{-\beta H_{\SS}^{\rm R}(t)}}{Z_{\SS}^{\rm R}(t)},
\end{equation}
with $Z_{\SS}^{\rm R}(t)=Z_{\SS}$ ($\Lambda_{t}^{\rm R}$ is trace preserving) and
\begin{align}\label{Href}
H_{\SS}^{\rm R}(t):=-\beta^{-1} \log \Lambda_{t}^{\rm R}\big(\ee^{-\beta H_{\SS}}\big).
\end{align}
Then, we introduce the \emph{refined weak coupling internal energy} by analogy to \eqref{EintWeak} as
\begin{equation}\label{EintRef}
E^{\rm R}(t):=\Tr[\rho_{\SS}(t)H_{\SS}^{\rm R}(t)].
\end{equation}
This definition differs from the standard weak coupling energy $E(t)$ \eqref{EintWeak} for finite times, but it presents the required limiting behavior:
\begin{enumerate}
\item $E^{\rm R}(0)=E(0)$. Thus, the deviation of $E^{\rm R}(t)$ from $E(t)$ at finite times is unambiguously caused by the interaction term $V_{\SE}$ in the Hamiltonian.
\bigskip
\item $E^{\rm R}(t)$ approaches $E(t)$ for $t$ large. In such a case $\Lambda_{t}^{\rm R}$ approaches the Davies' semigroup \eqref{REFapproxDavies} and then, $H_{\SS}^{\rm R}(t)$ approaches $H_{\SS}$. 
\end{enumerate}
Given that no work is considered, the first law reads
\begin{equation}\label{1lawRef0}
Q^{\rm R}(t):=\Delta E^{\rm R}(t)=E^{\rm R}(t)-E^{\rm R}(0),
\end{equation}
for the \emph{refined weak coupling heat}, or alternatively
\begin{equation}\label{RefQnoWork}
Q^{\rm R}(t):=\int_0^t \dot{E}^{\rm R}(r)dr= \int_0^t \Tr[\dot{\rho}_{\SS}(r)H_{\SS}^{\rm R}(r)]+\Tr[\rho_{\SS}(r)\dot{H}_{\SS}^{\rm R}(r)] dr,
\end{equation}
with 
\begin{equation}\label{RefQflowNoWork}
\dot{Q}^{\rm R}(t):=\Tr[\dot{\rho}_{\SS}(t)H_{\SS}^{\rm R}(t)]+\Tr[\rho_{\SS}(t)\dot{H}_{\SS}^{\rm R}(t)]
\end{equation}
the \emph{refined weak coupling heat flow}.

The second law can be derived in the integrated form. Since $\Lambda_{t}^{\rm R}$ is trace preserving and CP for any $t$, the monotonicity of the relative entropy \eqref{monotonicity} gives
\begin{equation}
S\Big\{\Lambda_{t}^{\rm R}[\rho_{\SS}(0)]\Big\Vert\Lambda_{t}^{\rm R}\big(\rho^{\beta}_{\SS}\big)\Big\}\leq S\Big[\rho_{\SS}(0)\Big\Vert \rho^{\beta}_{\SS}\Big],
\end{equation}
which can be straightforwardly recast in the form
\begin{equation}\label{2lawRefined}
\Delta S(t)-\beta Q^{\rm R}(t)\geq 0.
\end{equation}
The equality is reached for $\rho_{\SS}(0)=\rho^{\beta}_{\SS}$. This is a property shared with the standard weak coupling limit, but, as a key difference, $\rho_{\SS}^{\beta}$ is not invariant for the refined weak coupling [see Eqs. \eqref{RWCinfinityNEQ} and \eqref{RWCinfinity}].

Therefore, the present choice of internal energy is thermodynamically consistent for a system coupled with a thermal reservoir in the refined weak coupling limit (in absence of work). Note that, as aforementioned, since $\Lambda^{\rm R}_{t}$ is in general not P-divisible, a differential form for the second law as in \eqref{2lawweak} cannot be expected to be true in the refined case for finite times.

\subsection{Time-Dependent System Hamiltonian $H_{\SS}(t)$}
The situation becomes considerably more intricate if we allow for a time-dependent system Hamiltonian $H_{\SS}(t)$. This is so even for slowly-varying $H_{\SS}(t)$ such that the dynamics is well approximated by the adiabatically-deformed refined weak coupling method, $\Lambda_t^{\rm RA}$, Eqs. \eqref{LambdaRA1} and \eqref{LambdaRA2}. 

In analogy to \eqref{EintWeak} and \eqref{EintRef}, we can define internal energy by an equation of the form
\begin{equation}
E^{\rm RA}(t):=\Tr[\rho_\SS(t)H_{\SS}^{\rm RA}(t)],
\end{equation} 
with some appropriate choice of Hermitian operator $H_{\SS}^{\rm RA}(t)$ satisfying $H_{\SS}^{\rm RA}(0)=H_{\SS}(0)$, and $H_{\SS}^{\rm RA}(t)=H_{\SS}^{\rm R}(t)$ in \eqref{Href} for a time-independent $H_{\SS}$. However, the straightforward generalization of \eqref{Href} with the changes $\Lambda_t^{\rm R}\to \Lambda_t^{\rm RA}$ and $H_{\SS}\to H_{\SS}(t)$ is problematic. We should notice that the refined heat flow in \eqref{RefQflowNoWork} has a term with the time derivative $\dot{H}_{\SS}^{\rm R}(t)$, which is not zero because the dynamics is not given by a semigroup in the refined weak coupling limit. Hence, the choice of $H_{\SS}^{\rm RA}(t)$ is a subtle point, because the term $\Tr[\rho_{\SS}(t)\dot{H}_{\SS}^{\rm RA}(t)]$ in the time derivative of the internal energy $E^{\rm RA}(t)$ cannot be unambiguously identified with work. 

In order to find a suitable definition work, we may argue that the composition of the system and reservoir forms a closed system, so any energy change in this global system can be unambiguously identified with work. Thus, for $H(t)=H_{\SS}(t)+H_\EE+V_{\SE}$, using the von Neumann equation for $\dot{\rho}_{\SE}(t)$, we obtain
\begin{equation}
\frac{d}{dt}\langle H(t)\rangle=\Tr[\dot{\rho}_{\SE}(t)H(t)]+\Tr[\rho_{\SE}(t)\dot{H}(t)]=\Tr[\rho_{\SS}(t)\dot{H}_{\SS}(t)],
\end{equation}  
for the power applied to/performed by the global system. However, since it acts only on $\SS$ via the variation of $H_{\SS}(t)$, we adopt it as the definition of power for $\SS$ \cite{Esposito,Modi,Tanimura,StrasbergPRX,StrasbergPRE} also in the refined weak coupling limit:
\begin{equation}\label{WRAPower}
\dot{W}^{\rm RA}(t):=\dot{W}(t)=\Tr[\rho_{\SS}(t)\dot{H}_{\SS}(t)].
\end{equation}
Having in mind this definition of work, a convenient definition of $H_{\SS}^{\rm RA}(t)$ turns out to be: 
\begin{equation}\label{HRA}
H_{\SS}^{\rm RA}(t):=-\beta^{-1} \log \left[\Lambda_{t}^{\rm RA}\left\{\ee^{-\beta H_{\SS}(0)-\beta\int_0^t (\Lambda_{s}^{\rm RA})^\star[\dot{H}_{\SS}(s)]ds}\right\}\right],
\end{equation}
where $\Lambda^{\star}$ denotes the Heisenberg adjoint of $\Lambda$, $\Tr[\Lambda(A)B]=\Tr[A\Lambda^\star(B)]$. This is a Hermitian operator which satisfies the requirements $H_{\SS}^{\rm RA}(0)=H_{\SS}(0)$, and for time-independent $H_{\SS}$, $H_{\SS}^{\rm RA}(t)=H_{\SS}^{\rm R}(t)$, as we desired.

Thus, we state the \emph{first law} in the form
\begin{equation}\label{1lawRef}
\dot{E}^{\rm RA}(t)=\dot{Q}^{\rm RA}(t)+\dot{W}(t),
\end{equation}
with \emph{refined weak coupling heat flow} given by
\begin{equation}\label{RefQWork}
\dot{Q}^{\rm RA}(t):=\Tr[\dot{\rho}_{\SS}(t)H_{\SS}^{\rm RA}(t)]+\Tr[\rho_{\SS}(t)\dot{H}_{\SS}^{\rm RA}(t)]-\Tr[\rho_{\SS}(t)\dot{H}_{\SS}(t)].
\end{equation}
This way, for time-independent $H_{\SS}$,  $\dot{Q}^{\rm RA}(t)=\dot{Q}^{\rm R}(t)$ in \eqref{RefQflowNoWork}, and the first law \eqref{1lawRef} reduces to the case of the previous section \eqref{1lawRef0}. For time-dependent $H_{\SS}(t)$, the last term in \eqref{RefQWork} aims to subtract the direct dependence of the first two terms with the time derivative $\dot{H}_{\SS}(t)$ which we have considered to define work, Eq. \eqref{WRAPower}. Hence, the integrated heat is
\begin{equation}
Q^{\rm RA}(t)=\Tr[\rho_{\SS}(t)H^{\rm RA}_{\SS}(t)]-\Tr[\rho_{\SS}(0)H_{\SS}(0)]-\int_0^t\Tr[\rho_{\SS}(s)\dot{H}_{\SS}(s)]ds.
\end{equation}
In order to derive the second law, we define the auxiliary object
\begin{equation}
\Omega(t,r):=-\beta^{-1} \log \left[\Lambda_{t}^{\rm RA}\left\{\ee^{-\beta H_{\SS}(0)-\beta\int_0^r (\Lambda_{s}^{\rm RA})^\star[\dot{H}_{\SS}(s)]ds}\right\}\right],
\end{equation}
which satisfies $\Omega(t,t)=H_{\SS}^{\rm RA}(t)$. A straightforward computation gives the heat written in terms of $\Omega(t,r)$ as
\begin{equation}\label{RefQaux}
Q^{\rm RA}(t)=\Tr[\rho_{\SS}(t)\Omega(t,t)]-\Tr[\rho_{\SS}(0)\Omega(0,t)].
\end{equation}
For the state
\begin{equation}
\rho_{0}(\beta,r):=Z^{-1}_\SS(r)\ee^{-\beta H_{\SS}(0)-\beta\int_0^r (\Lambda_{s}^{\rm RA})^\star[\dot{H}_{\SS}(s)]ds},
\end{equation}
the monotonicity of the relative entropy \eqref{monotonicity} gives
\begin{equation}
S\Big\{\Lambda_{t}^{\rm RA}[\rho_{\SS}(0)]\Big\Vert\Lambda_{t}^{\rm RA}[\rho_{0}(\beta,r)]\Big\}\leq S\Big[\rho_{\SS}(0)\Big\Vert \rho_{0}(\beta,r)\Big],
\end{equation}
which can be recast in the form
\begin{equation}
\Delta S(t)-\beta\left\{\Tr[\rho_{\SS}(t)\Omega(t,r)]-\Tr[\rho_{\SS}(0)\Omega(0,r)]\right\}\geq 0.
\end{equation}
Since this is fulfilled for all $r$, and particularly for $r=t$, according to \eqref{RefQaux} we obtain the \emph{second law}:
\begin{equation}
\Delta S(t)-\beta Q^{\rm RA}(t)\geq0.
\end{equation}
This completes the thermodynamic formulation of the refined weak coupling limit.

On the other hand, it is worth noticing that the refined internal energy $\dot{E}^{\rm RA}(t)$ approaches $\dot{E}(t)$ at large times for time-dependent $H_{\SS}(t)$ in the very slow variation limit where Eq. \eqref{Ergodict} holds true. However, in this limit, the work is actually approaching zero in the time scale $t\lesssim \alpha^{-2}$, where the refined and the standard weak coupling limit are good descriptions of the exact dynamics. This manifests that the asymptotic thermodynamic behavior of the refined weak coupling limit  for slowly-varying $H_{\SS}(t)$ is a nontrivial point.

\section{Example: Spin-Boson model in the Refined Weak Coupling Limit}
As an example of the previous results, we can consider the (transverse) spin-boson problem studied in \cite{Refined} in the refined weak coupling limit. Consider first the case of a time-independent system Hamiltonian,
\begin{equation}
H=\frac{\omega_0}{2}\sigma_z+\sum_{k} \omega_k a_k^\dagger a_k+\sum_{k} \sigma_x g_k(a_k+a_k^\dagger),
\end{equation}
with $H_{\SS}=\frac{\omega_0}{2}\sigma_z$, $H_{\EE}= \sum_{k} \omega_k a_k^\dagger a_k$, and $V_{\SE}=\sum_{k} \sigma_x g_k(a_k+a_k^\dagger)$. The exponent $\mathcal{Z}_2(t)$ is obtained in the form \cite{Refined}:
\begin{align}\label{Zsb}
\mathcal{Z}_2(t)(\rho_\SS)=&-i[\Xi(t,\beta)\sigma_z,\rho_\SS]+\sum_{\mu,\nu=+,-}\Gamma_{\mu\nu}(t,\beta)[\sigma_\nu\rho_\SS\sigma_\mu^\dagger-\{\sigma_\mu^\dagger\sigma_\nu,\rho_\SS\}].
\end{align}
Here, $\sigma_\pm=(\sigma_x\pm i\sigma_y)/2$ are the lowering and raising Pauli matrices and the coefficients are given by
\begin{align}
&\Xi(t,\beta)=\frac{1}{4\pi}\int_{-\infty}^\infty d\omega t^2\left\{\mathrm{sinc}^2\left[\tfrac{(\omega_0-\omega)t}{2}\right]-\mathrm{sinc}^2\left[\tfrac{(\omega_0+\omega)t}{2}\right]\right\}\left\{\mathrm{P.V.}\int_0^\infty d\upsilon J(\upsilon)\left[\tfrac{\bar{n}_\beta(\upsilon)+1}{\omega-\upsilon}+\tfrac{\bar{n}_\beta(\upsilon)}{\omega+\upsilon}\right]\right\},\\
&\Gamma_{--}(t,\beta)=\int_0^\infty d\omega t^2J(\omega)\left\{[\bar{n}_\beta(\omega)+1]\mathrm{sinc}^2\left[\tfrac{(\omega_0-\omega)t}{2}\right]+\bar{n}_\beta(\omega)\mathrm{sinc}^2\left[\tfrac{(\omega_0+\omega)t}{2}\right]\right\},\\
&\Gamma_{++}(t,\beta)=\int_0^\infty d\omega t^2J(\omega)\left\{[\bar{n}_\beta(\omega)+1]\mathrm{sinc}^2\left[\tfrac{(\omega_0+\omega)t}{2}\right]+\bar{n}_\beta(\omega)\mathrm{sinc}^2\left[\tfrac{(\omega_0-\omega)t}{2}\right]\right\},\\
&\Gamma_{+-}(t,\beta)=\Gamma_{-+}^\ast(t,\beta)=\int_0^\infty d\omega t^2J(\omega)[2\bar{n}_\beta(\omega)+1]e^{-i\omega_0t}\mathrm{sinc}\left[\tfrac{(\omega_0+\omega)t}{2}\right]\mathrm{sinc}\left[\tfrac{(\omega_0-\omega)t}{2}\right],\label{ZsbCoef}
\end{align}
where $J(\omega)$ is the spectral density of the bath, $\bar{n}_\beta(\omega)=[\exp(\beta \omega)-1]^{-1}$ is the mean number of bosons in the bath with frequency $\omega$, and $\mathrm{sinc}(\omega):=\tfrac{\sin \omega}{\omega}$.

\begin{figure}[t!]
\centering
\includegraphics[width=0.8\textwidth]{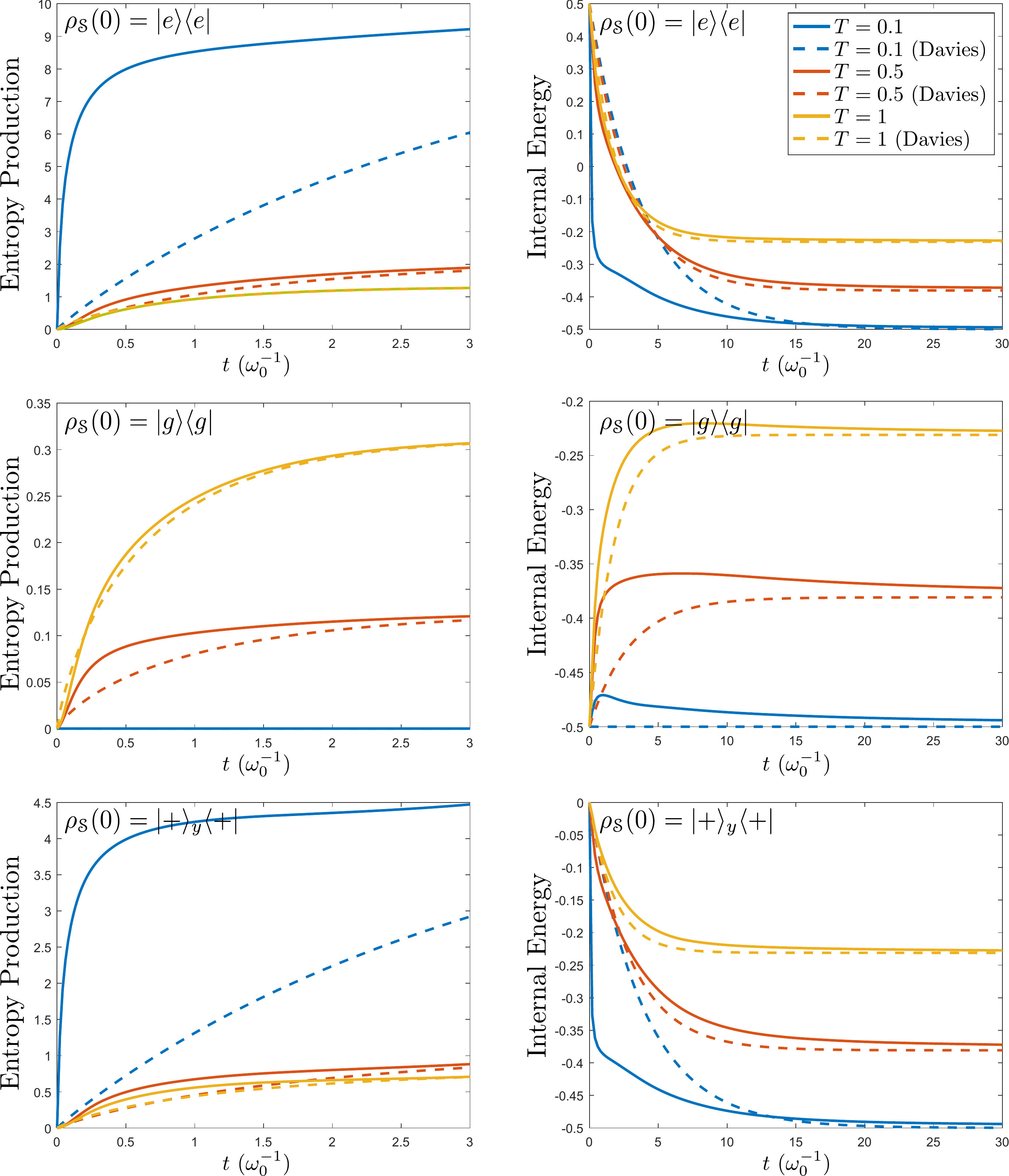}
\caption{Results for the entropy production (left column) and internal energy (right column) for the spin-boson model under the refined weak coupling limit (solid lines) and the Davies semigroup dynamics of the weak coupling limit (same color, dashed lines). These are calculated under three different system initial conditions $\rho_{\SS}(0)=|e\rangle\langle e|$, $\rho_{\SS}(0)=|g\rangle\langle g|$, and $\rho_{\SS}(0)=|+\rangle\mbox{}_y\langle +|$, which are depicted in the first, second, and third row, respectively. The bath is assumed to have an Ohmic spectral density $J(\omega)=\alpha \omega \exp(-\omega/\omega_c)$ with $\alpha=0.05$ and $\omega_c=5$, in units of $\omega_0$. The different bath temperatures are highlighted by different colors. As expected, convergence for large time is obtained.}
\end{figure}  

We can then compute the refined weak coupling internal energy. In this case, the map $\Lambda^{\rm R}_t=\ee^{-\ii \HH_z t}\ee^{\mathcal{Z}_2(t)}$  is an incoherent map in the $\sigma_z$ eigenbasis ($\HH_z:=\tfrac{\omega_0}{2}[\sigma_z,\cdot]$), so that the operator $H_\SS^{\rm R}$ in \eqref{Href} remains diagonal in the same basis as $H_{\SS}$:
\begin{equation}
H_\SS^{\rm R}(t)=-\beta^{-1}\begin{pmatrix}h_{11}(t) & 0\\
0 & h_{22}(t)
\end{pmatrix}.
\end{equation}
A straightforward computation yields
\begin{align}
h_{11}(t):=\log \frac{2\Gamma_{++}(t,\beta)\mathrm{cosh}\left(\tfrac{\omega_0 \beta }{2}\right)-\ee^{-[\Gamma_{++}(t,\beta)+\Gamma_{--}(t,\beta)]}\left[\ee^{\tfrac{\omega_0 \beta }{2}}\Gamma_{++}(t,\beta)-\ee^{-\tfrac{\omega_0 \beta }{2}} \Gamma_{--}(t,\beta)\right] }{\Gamma_{++}(t,\beta)+\Gamma_{--}(t,\beta)},
\end{align}
\begin{align}
h_{22}(t):=\log \frac{2\Gamma_{--}(t,\beta)\mathrm{cosh}\left(\tfrac{\omega_0 \beta }{2}\right)+\ee^{-[\Gamma_{++}(t,\beta)+\Gamma_{--}(t,\beta)]}\left[\ee^{\tfrac{\omega_0 \beta }{2}}\Gamma_{++}(t,\beta)-\ee^{-\tfrac{\omega_0 \beta }{2}} \Gamma_{--}(t,\beta)\right] }{\Gamma_{++}(t,\beta)+\Gamma_{--}(t,\beta)}.
\end{align}
The non-Markovianity of this model has been studied in \cite{Refined}, showing that it breaks P-divisibility and it is ``quasieternal'' CP-indivisible. 

We have examined the entropy production \eqref{2lawRefined} and the refined internal energy \eqref{EintRef} for three different initial states, the excited $|e\rangle$  and ground $|g\rangle$ states, and the superposition state $|+\rangle_y$, $\sigma_y|+\rangle_y=|+\rangle_y$. The results are shown in Fig. 1 for three different temperatures, jointly with the results obtained for the Davies quantum dynamical semigroup in the standard weak coupling limit. We find the expected convergence towards the internal energy and the entropy production of the weak coupling. Despite the model is not P-divisible, we do not see oscillations in the entropy production. This is probably caused by the diagonal form of the operator $H_{\SS}^{\rm R}(t)$. These results seem to indicate that the larger difference between the standard weak coupling limit and the refined one arises at low bath temperatures. This is in agreement of what was obtained in \cite{Refined}, and somehow expected because the width of the reservoir correlation functions increases when $T$ decreases \cite{MME}.

\begin{figure}[t!]
\centering
\includegraphics[width=0.9\textwidth]{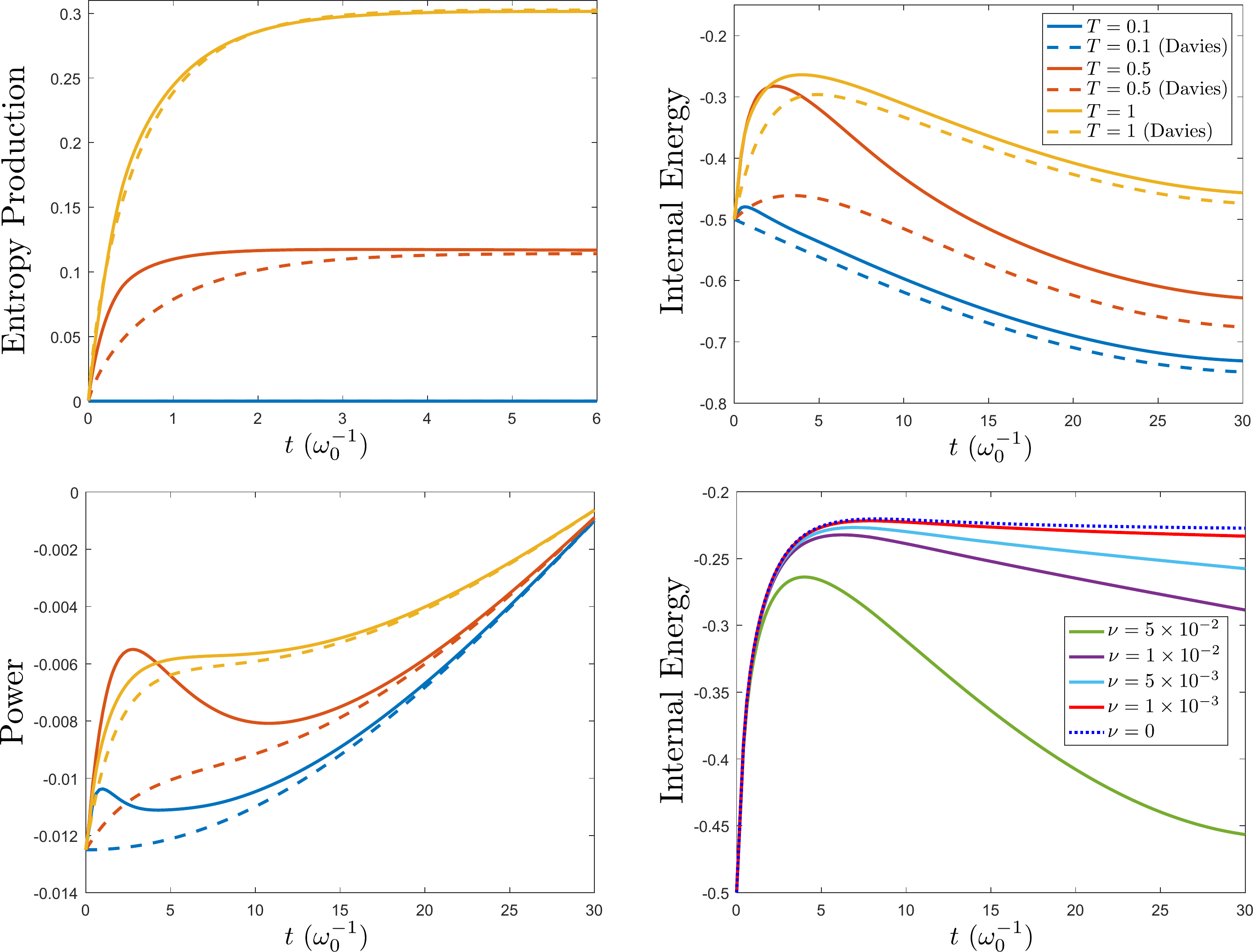}
\caption{Results for the spin-boson model with diagonal driving $H_\SS(t)=\tfrac{\omega_0(t)}{2}\sigma_z$ under the adiabatically-deformed refined and Davies weak coupling limit (same color, dashed lines). The entropy production (top left column), power (bottom left column), and internal energy (top right column) are plotted for three different bath temperatures. The internal energy for different values of the modulation frequency at $T=1$ is also depicted (bottom right column) showing convergence to the refined weak coupling result for constant $H_\SS$ ($\nu=0$, blue dotted line). These results are calculated for the system initially prepared in the ground state $\rho_{\SS}(0)=|g\rangle\langle g|$. As in Fig. 2, the bath is assumed to have an Ohmic spectral density $J(\omega)=\alpha \omega \exp(-\omega/\omega_c)$ with $\alpha=0.05$ and $\omega_c=5$, in units of $\omega_0$.}
\end{figure}

As a further example including a system time-dependent Hamiltonian, we can consider the same model with a diagonal driving given by the frequency modulation:
\begin{equation}
H_{\SS}(t)=\frac{\omega_0(t)}{2}\sigma_z, \quad \omega_0(t)=\omega_0(0)+\lambda \sin(\nu t).
\end{equation}  
For $\nu$ small enough, the dynamics can be approximated by the adiabatically-deformed refined weak coupling limit $\Lambda^{\rm RA}_t$, Eqs. \eqref{LambdaRA1} and \eqref{LambdaRA2}. Thus, the structure of the dynamics is the same as before with the generator $\mathcal{Z}_2(t,s)$ as in Eqs. \eqref{Zsb}--\eqref{ZsbCoef}, but $\omega_0(s)$ in the place of the formerly constant $\omega_0$. 

Despite involving an intricate time-ordering operation, if~the system is initially in a diagonal state in the $\sigma_z$-eigenbasis, e.g.,~the~ground state $|g\rangle\langle g|$, the~calculation of $\Lambda^{\rm RA}_t(|g\rangle\langle g|)$ can be simplified by a series of algebraic manipulations explained in detail in Appendix \ref{B}. The same methods can be applied in the computation of the operator $H^{\rm RA}_{\SS}(t)$, Eq. \eqref{HRA}, which also remains diagonal for all $t$. This was done analytically up to the numerical computation of two (standard) integrals. The results are depicted in Fig. 2 for several temperatures and modulation frequencies.

Since $\Lambda^{\rm RA}_t$ approaches \eqref{DaviesAd} for large times, convergence to the adiabatically-deformed weak coupling limit was found for the power \eqref{WRAPower}. For the internal energy, there is a $\nu$-decreasing small mismatch between the internal energy given by both techniques even at large time. As aforementioned, the full equivalence is found for very small $\nu$. The entropy production is positive and again its rate of change is also positive. This was something expected after the results for the time-independent case and the fact that the modulation keeps $H_{\SS}(t)$ diagonal in the same basis. Finally as $\nu$ approaches zero we obtain consistency with the internal energy for the refined weak coupling with a time-independent system Hamiltonian.

\section{Conclusions}
In this contribution, we have formulated a thermodynamic framework for the refined weak coupling limit, which represents a non-Markovian approach to the evolution of open quantum systems. We have included the case of time-independent system Hamiltonians, as well as slowly-varying time-dependent ones. To this end, we have extended the refined weak coupling techniques to deal with slowly-varying time-dependent Hamiltonians and redefine the internal energy because of the non-negligible effect of the system-environment interaction. In the appropriate limiting situation, this refined internal energy approaches the usual internal energy in the weak coupling. Finally, we have illustrated the results by analyzing the refined internal energy and the entropy production of a two-level system in contact with a thermal bath. In this example, both operators $H^{\rm R}_{\SS}(t)$ and $H^{\rm RA}_{\SS}(t)$ are diagonal in the same basis as $H_{\SS}$, which is probably a reason way we see not only positive entropy production but also a positive entropy production rate for the initial states analyzed. 

As a future work, several extensions of this approach may be considered. For instance: the cases of periodic time-dependent $H_\SS(t)$ under the Floquet formalism \cite{AlickiTutorialPeriodic}, and systems coupled with several heat baths. Furthermore, since some of the arguments employed in Sec. 4 only rely on the fact that $\Lambda_t^{\rm R}$ and $\Lambda_t^{\rm RA}$ are CP dynamical maps, one may wonder if the same proposal can be used to formulate a thermodynamic framework for any CP dynamical map. This seems an interesting possibility to be analyzed in the near future.

\vspace{6pt}

\acknowledgments{The author acknowledges the Spanish MINECO grants FIS2015-67411, FIS2017- 91460-EXP, the CAM research consortium QUITEMAD S2018/TCS-4342, and US Army Research Office through grant W911NF-14-1-0103 for partial financial support. He is grateful to Aurelia Chenu and Adolfo del Campo for sharing their insight on this subject during his visit to DIPC. He also thanks them for their kind invitation and hospitality.}

\conflictsofinterest{The author declares no conflict of interest.} 



\appendixtitles{yes} 
\appendix
\section{Internal Energy in the Weak Coupling Limit\label{A}}
Consider the projection operators $\mathcal{P}\rho_{\SE}=\Tr_{\EE}(\rho_{\SS})\otimes \rho_{\EE}^\beta$, and $\mathcal{Q}=\mathds{1}-\mathcal{P}$. By applying $\mathcal{Q}$ on the the von Neumann equation \eqref{VNeq}: 
\begin{equation}
\frac{d}{dt}\mathcal{Q}\tilde{\rho}_{\mathscr{SE}}(t)=\mathcal{Q}\tilde{\mathcal{V}}(t)\tilde{\rho}_{\mathscr{SE}}(t)=\mathcal{Q}\tilde{\mathcal{V}}(t)\mathcal{P}\tilde{\rho}_{\mathscr{SE}}(t)+\mathcal{Q}\tilde{\mathcal{V}}(t)\mathcal{Q}\tilde{\rho}_{\mathscr{SE}}(t),
\end{equation}
where we have inserted the identity between $\tilde{\mathcal{V}}(t)$ and $\tilde{\rho}_{\mathscr{SE}}(t)$. The formal integration of this equation gives \cite{BrPr02,Libro}:
\begin{equation}
\mathcal{Q}\tilde{\rho}_{\mathscr{SE}}(t)=\mathcal{G}(t,0)\mathcal{Q}\rho_{\mathscr{SE}}(0)+\int_0^tds\mathcal{G}(t,s)\mathcal{Q}\tilde{\mathcal{V}}(s)\mathcal{P}\tilde{\rho}_{\mathscr{SE}}(s),
\end{equation}
with
\begin{equation}
\mathcal{G}(t,s)=\mathcal{T}\ee^{\int_s^tdr\mathcal{Q\tilde{V}}(r)}.
\end{equation}
For $\rho_{\SE}(0)=\rho_{\SS}(0)\otimes\rho_{\EE}^\beta$, $\mathcal{Q}\rho_{\mathscr{SE}}(0)=0$ and the first term of the above equation vanishes:
\begin{equation}\label{appQ}
\mathcal{Q}\tilde{\rho}_{\mathscr{SE}}(t)=\int_0^tds\mathcal{G}(t,s)\mathcal{Q}\tilde{\mathcal{V}}(s)\mathcal{P}\tilde{\rho}_{\mathscr{SE}}(s).
\end{equation}
On the other hand, in the weak coupling limit:
\begin{equation}\label{auxPweak}
\mathcal{P}\tilde{\rho}_{\SE}(t)=\tilde{\rho}_\SS(t)\otimes\ \rho_{\EE}^\beta=\ee^{\alpha^2\tilde{\mathcal{L}}t}\rho_\SS(0)\otimes\ \rho_{\EE}^\beta=\ee^{\alpha^2\tilde{\mathcal{L}}t}\otimes\mathds{1}[\mathcal{P}\rho_{\SE}(0)].
\end{equation}
By introducing the projection operators in the mean value of $H_\SS+\alpha V_{\SE}$ (in the interaction picture), we obtain
\begin{align}\label{auxEint}
\langle [H_\SS+\alpha V_{\SE}](t)\rangle&=\Tr\{[\tilde{H}_\SS(t)+\alpha\tilde{V}_{\SE}(t)]\tilde{\rho}_{\SE}(t)\}\nonumber\\
&=\Tr\{[\tilde{H}_\SS(t)+\alpha\tilde{V}_{\SE}(t)]\mathcal{P}\tilde{\rho}_{\SE}(t)\}+\Tr\{[\tilde{H}_\SS(t)+\alpha\tilde{V}_{\SE}(t)]\mathcal{Q}\tilde{\rho}_{\SE}(t)\}.
\end{align}
The last term in this equation is simplified because $\Tr[A\mathcal{Q}\tilde{\rho}_{\SE}(t)]=0$ for any system operator $A$ and using \eqref{appQ}:
\begin{equation}
\Tr\{[\alpha\tilde{V}_{\SE}(t)]\mathcal{Q}\tilde{\rho}_{\SE}(t)\}= \alpha^2\int_0^tds\Tr[\tilde{V}_{\SE}(t)\mathcal{G}(t,s)\mathcal{Q}\tilde{\mathcal{V}}(s)\mathcal{P}\tilde{\rho}_{\mathscr{SE}}(s)]=\alpha^2 \Tr[\mathcal{R}(t,s)\mathcal{P}\tilde{\rho}_{\mathscr{SE}}(t)], 
\end{equation}
where we have taken the coupling constant $\alpha$ outside of $\tilde{\mathcal{V}}(s)$, and after \eqref{auxPweak},
\begin{equation}
\mathcal{R}(t,s):=\tilde{V}_{\SE}(t)\int_0^tds\mathcal{G}(t,s)\mathcal{Q}\tilde{\mathcal{V}}(s) \left[\ee^{\alpha^2\tilde{\mathcal{L}}(s-t)}\otimes\mathds{1}\right].
\end{equation}
Furthermore, since $\Tr\big[\tilde{V}_{\SE}\rho_\SE^\beta\big]=0$, $\Tr\big[\tilde{V}_{\SE}(t)\mathcal{P}\tilde{\rho}_{\SE}(t)\big]=0$, and Eq. \eqref{auxEint} yields
\begin{equation}
\langle [H_\SS+\alpha V_{\SE}](t)\rangle=\Tr\{[\tilde{H}_\SS(t)+\alpha^2 R(t,s)]\mathcal{P}\tilde{\rho}_{\SE}(t)\}=\Tr[\tilde{H}_\SS(t)\tilde{\rho}_\SS(t)]+\mathcal{O}(\alpha^2),
\end{equation}
hence
\begin{equation}
\langle H(t)\rangle\simeq\Tr[H_\SS\rho_\SS(t)]+\Tr[H_\EE\rho_\EE(t)]+\mathcal{O}(\alpha^2).
\end{equation}

\setcounter{equation}{0}
\renewcommand*{\theequation}{B\arabic{equation}}

\section{Calculation of the Time-Ordered Exponential for the Driven Spin-Boson Model in the Refined Weak Coupling\label{B}}
For the spin-boson model with diagonal driving $H_{\SS}(t)=\tfrac{\omega_0(t)}{2}\sigma_z$, the operator $\mathcal{Z}_2(t,s)$ has the same algebraic structure as \eqref{Zsb}. It is a linear combination of the operators $\mathcal{Z}_z:=[\sigma_z,\cdot]$,  $\mathcal{D}_{+-}:=\sigma_+(\cdot)\sigma_--\{\sigma_-\sigma_+,\cdot\}/2$, $\mathcal{D}_{-+}:=\sigma_-(\cdot)\sigma_+-\{\sigma_+\sigma_-,\cdot\}/2$, $\mathcal{D}_{--} :=\sigma_-(\cdot)\sigma_-$, and $\mathcal{D}_{++}:=\sigma_+(\cdot)\sigma_+$. As commented in \cite{Refined}, these operators close a Lie algebra:
\begin{align}
&[\mathcal{D}_{+-},\mathcal{D}_{-+} ]=\mathcal{D}_{+-}-\mathcal{D}_{-+},\quad [\mathcal{D}_{++},\mathcal{D}_{--} ]=\Sigma_z/2, \nonumber\\
&[\Sigma_z,\mathcal{D}_{++} ]=4\mathcal{D}_{++},\quad [\Sigma_z,\mathcal{D}_{--} ]=-4\mathcal{D}_{--},
\end{align}
with the zero value for the rest of the cases. Actually, it is clear that this algebra is the direct sum of the subalgebras generated by $\{\mathcal{D}_{+-},\mathcal{D}_{-+}\}$ and $\{\mathcal{Z}_{z},\mathcal{D}_{++},\mathcal{D}_{--}\}$, so that the temporally-ordered exponential in \eqref{LambdaRA1} splits into
\begin{align}\label{AppB1}
\tilde{\Lambda}_t^{\rm RA}&=\mathcal{T} \ee^{\frac{1}{t}\int_0^{t}\mathcal{Z}_2(t,s) ds}\nonumber\\
&=\left\{\mathcal{T} \ee^{\frac{1}{t}\int_0^{t}\left\{\Gamma_{++}(t,s,\beta)\mathcal{D}_{+-}+\Gamma_{--}(t,s,\beta)\mathcal{D}_{-+}\right] ds}\right\}\left\{\mathcal{T} \ee^{\frac{1}{t}\int_0^{t}\left[-\ii\Xi(t,s,\beta)\Sigma_z+\Gamma_{+-}(t,s,\beta)\mathcal{D}_{--}+\Gamma_{-+}(t,s,\beta)\mathcal{D}_{++}\right] ds}\right\}.
\end{align} 
The second one leaves invariant any diagonal operator $D$ in the $\sigma_z$ eigenbasis, so that from \eqref{LambdaRA2}
\begin{equation}
\Lambda_t^{\rm RA}(D)=\ee^{-\tfrac{\ii}{2}\int_0^t \omega_0(s)\Sigma_z ds}\tilde{\Lambda}_t^{\rm RA}(D)=\mathcal{T} \ee^{\frac{1}{t}\int_0^{t}\left\{\Gamma_{++}(t,s,\beta)\mathcal{D}_{+-}+\Gamma_{--}(t,s,\beta)\mathcal{D}_{-+}\right] ds}(D).
\end{equation}
Therefore, the application of $\Lambda_t^{\rm RA}$ on diagonal states only requires the calculation of the first time-ordered exponential between curly brackets of \eqref{AppB1}. It is straightforward to prove that the same is true for the Heisenberg adjoint $(\Lambda_t^{\rm RA})^\star$ acting on diagonal operators.

In addition, the fact that $\{\mathcal{D}_{+-},\mathcal{D}_{-+}\}$ closes a Lie algebra allows for writing the required temporally ordered exponential as a product of exponentials of these operators \cite{Puri}, 
\begin{equation}\label{AppB2}
\mathcal{T} \ee^{\frac{1}{t}\int_0^{t}\left\{\Gamma_{++}(t,s,\beta)\mathcal{D}_{+-}+\Gamma_{--}(t,s,\beta)\mathcal{D}_{-+}\right] ds}=\ee^{C_{++}(t)\mathcal{D}_{+-}}\ee^{C_{--}(t)\mathcal{D}_{-+}}.
\end{equation} 
In order to compute the numbers $C_j(t)$, we introduce a new ones $C_j(t,r)$ for the decomposition
\begin{equation}\label{AppB3}
\mathcal{T} \ee^{\frac{1}{r}\int_0^{t}\left\{\Gamma_{++}(r,s,\beta)\mathcal{D}_{+-}+\Gamma_{--}(r,s,\beta)\mathcal{D}_{-+}\right] ds}=\ee^{C_{++}(t,r)\mathcal{D}_{+-}}\ee^{C_{--}(t,r)\mathcal{D}_{-+}},
\end{equation} 
such that $C_j(t,t)=C_j(t)$. Differentiating with respect to $t$ and multiplying by the inverse \eqref{AppB3} at both sides yields
\begin{align}\label{AppB4}
\tfrac{\Gamma_{++}(r,s,\beta)}{r}\mathcal{D}_{+-}+\tfrac{\Gamma_{--}(r,s,\beta)}{r}\mathcal{D}_{-+} &=\dot{C}_{++}(t,r)\mathcal{D}_{+-}+\dot{C}_{--}(t,r)\Big[\ee^{C_{++}(t,r)\mathcal{D}_{+-}}\Big]\mathcal{D}_{-+}\Big[\ee^{-C_{++}(t,r)\mathcal{D}_{+-}}\Big].
\end{align}
The computation of the exponential products on the right-hand side, and a comparison between the coefficients of $\mathcal{D}_{+-}$ and $\mathcal{D}_{-+}$ on both sides leads to a pair of first order non-linear differential equations,
\begin{equation}\begin{cases}
\frac{\Gamma_{++}(r,s,\beta)}{r}=\dot{C}_{++}(t,r)+\left[1-\ee^{-C_{++}(t,r)}\right]\dot{C}_{--}(t,r),\\
\frac{\Gamma_{--}(r,s,\beta)}{r}=\ee^{-C_{++}(t,r)}\dot{C}_{--}(t,r).
\end{cases}
\end{equation}
These equations are easy to solve. Under the initial conditions $C_j(0,r)=0$, we obtain the solution
\begin{align}
&C_{--}(t,r)=\log\left\{1+\frac{1}{r}\int_0^t \Gamma_{--}(r,s,\beta)\ee^{\tfrac{1}{r}\int_0^s \left[\Gamma_{++}(r,u,\beta)+\Gamma_{--}(r,u,\beta)\right] du} ds \right\},\\
&C_{++}(t,r)=\frac{1}{r}\int_0^t \left[\Gamma_{++}(r,s,\beta)+\Gamma_{--}(r,s,\beta)\right] ds-C_{--}(t,r).
\end{align}
Finally, the coefficients $C_j(t)$ are
\begin{align}
&C_{--}(t)=\log\left\{1+\frac{1}{t}\int_0^t \Gamma_{--}(t,s,\beta)\ee^{\tfrac{1}{t}\int_0^s \left[\Gamma_{++}(t,u,\beta)+\Gamma_{--}(t,u,\beta)\right] du} ds \right\},\\
&C_{++}(t)=\frac{1}{t}\int_0^t \left[\Gamma_{++}(t,s,\beta)+\Gamma_{--}(t,s,\beta)\right] ds-C_{--}(t),
\end{align}
which allows us to compute the time-ordered exponential at \eqref{AppB2} by composing ordinary matrix exponentials.

\reftitle{References}


\begin{thebibliography}{999}
\bibitem{BrPr02}
Breuer, H.-P.; Petruccione, F. \textit{The Theory of Open Quantum Systems}; Oxford University Press: Oxford, UK, 2002.
%
\bibitem{GardinerZoller04} Gardiner, C. W.; Zoller, P. \textit{Quantum Noise}; Springer: Berlin, Germany, 2004.
%
\bibitem{Libro}
Rivas, A.; Huelga, S.~F. {\em Open Quantum Systems. An Introduction}; Springer: Heidelberg, Germany, 2012.
%
\bibitem{Qthermo1}
Gemmer, J.; Michel, M.; Mahler, G. {\em Quantum thermodynamics: Emergence of thermodynamic behavior within composite quantum systems}; Springer: Berlin, Germany, 2004.
%
\bibitem{QthermoKosloff} Kosloff, R. Quantum Thermodynamics: A Dynamical Viewpoint. {\em Entropy} {\bf 2013}, {\em 15}, 2100. 
%
\bibitem{Qthermo2} Binder, F.; Correa, L.A.; Gogolin, C.; Anders, J.; Adesso, G., Eds. {\em Thermodynamics in the Quantum Regime}; Springer: Cham, Switzerland, 2018. 
%
\bibitem{Davies} Davies, E.~B.; Markovian Master Equations. {\em Comm. Math. Phys.} {\bf 1974}, {\em 39}, 91.
%
\bibitem{GKLS1} Gorini, V.; Kossakowski, A.; Sudarshan, E.C.G. Completely positive dynamical semigroups of
N-level systems. {\em J. Math. Phys.} {\bf 1976}, {\em 17}, 821.
 Lindblad, G. On the generators of quantum dynamical semigroups. {\em Comm. Math. Phys.} {\bf 1976},
{\em 48}, 119.
%
\bibitem{GKLS2} Chru\'sci\'nski, D.; Pascazio, S. A Brief History of the GKLS Equation. {\em Open Sys. Inf. Dyn.} {\bf 2017}, {\em 24}, 1740001.
%
\bibitem{AlickiBook} Alicki, R.; Lendi, K. {\em Quantum Dynamical Semigroups and Applications}; Springer: Berlin,
Germany, 1987.
%
\bibitem{SpohnEntropy} Spohn, H. Entropy production for quantum dynamical semigroups. {\em J. Math. Phys.} {\em 1978}, {\em 19}, 1227.
%
\bibitem{SpohnLebowitz} Spohn, H.; Lebowitz, J. Irreversible thermodynamics for quantum systems weakly coupled to
thermal reservoirs. {\em Adv. Chem. Phys.} {\bf 1979}, {\em 38}, 109.
%
\bibitem{AlickiHeatEngine} Alicki, R. Quantum open systems as a model of a heat engine. {\em J. Phys A: Math. Gen.} {\bf 1979}, {\em 12}, L103.
%
\bibitem{AlickiTutorialPeriodic} Alicki, R.; Gelbwaser-Klimovsky, D.; Kurizki, G. Periodically driven quantum open systems: Tutorial. {\bf 2012}, arXiv:1205.4552.
%
\bibitem{Esposito} Esposito, M.; Lindberg, K.; van den Broek, C.; Entropy production as correlation between system and reservoir. {\em New J. Phys.} {\bf 2010}, {\em 12}, 013013.
%
\bibitem{Modi} Binder, F.~C.; Vinjanampathy, S.; Modi, K.; Goold, J. Quantum thermodynamics of general quantum processes. {\em Phys. Rev. E} {\bf 2015}, {\em 91}, 032119.
%
\bibitem{Tanimura} Kato, A.; Tanimura, Y. Quantum heat current under non-perturbative and non-Markovian conditions: Applications to heat machines. {\em J. Chem. Phys.} {\bf 2016}, {\em 145}, 224105.
%
\bibitem{Alipour1} Alipour, S.; Benatti, F.; Bakhshinezhad, F.; Afsary, M.; Marcantoni, S.; Rezakhani, A.~T. Correlations in quantum thermodynamics. {\em Sci. Rep.} {\bf 2016}, {\em 6}, 35568.
%
\bibitem{Alipour2} Marcantoni, S.; Alipour, S.; Benatti, F.; Floreanini, R.; Rezakhani, A.~T. Entropy production and non-Markovian dynamical maps. {\em Sci. Rep.} {\bf 2017}, {\em 7}, 12447.
%
\bibitem{StrasbergPRX} Strasberg, P.; Schaller, G.; Brandes, T.; Esposito, M. Quantum and Information Thermodynamics: A Unifying Framework Based on Repeated Interactions. {\em Phys. Rev. X} {\bf 2017}, {\em 7}, 021003.
%
\bibitem{Pati} Bhattacharya, S.; Misra, A.; Mukhopadhyay, C.; Pati, A.~K. Exact master equation for a spin interacting with a spin bath: Non-Markovianity and negative entropy production rate. {\em Phys. Rev. A} {\bf 2017}, {\em 95}, 012122.
%
\bibitem{Ghosh} Thomas, G.; Siddharth, H.; Banerjee, S.; Ghosh, S. Thermodynamics of non-Markovian reservoirs and heat engines. {\em Phys. Rev. E} {\bf 2018}, {\em 97}, 062108.
%
\bibitem{StrasbergPRE} Strasberg, P.; Esposito, M. Non-Markovianity and negative entropy production rates. {\em Phys. Rev. E} {\bf 2019}, {\em 99}, 012120.
%
\bibitem{Gelin} Gelin, M.~F.; Thoss, M. Thermodynamics of a subensemble of a canonical ensemble. {\em Phys. Rev. E} {\bf 2009}, {\em 79}, 051121.
%
\bibitem{Seifert} Seifert, U. First and second law of thermodynamics at strong coupling. {\em Phys. Rev. Lett.} {\bf 2016}, {\em 116}, 020601.
%
\bibitem{Hu} Hsiang, J.-T.; Hu, B,-L. Quantum Thermodynamics at Strong Coupling: Operator Thermodynamic Functions and Relations. {\em Entropy} {\bf 2018}, {\em 20}, 423. 
%
\bibitem{AlickiRefined} Alicki, R. Master equations for a damped nonlinear oscillator and the validity
of the Markovian approximation. {\em Phys. Rev. A} {\bf 1989}, {\em 40}, 4077.
%
\bibitem{SchallerBrandes} Schaller, G; Brandes, T. Preservation of positivity by dynamical coarse graining. {\em  Phys. Rev. A} {\bf 2008}, {\em 78}, 022106.
%
\bibitem{Benatti1} Benatti, F.; Floreanini, R.; Marzolino, U. Environment induced entanglement in a refined weak-coupling limit. {\em EPL} {\bf 2009}, {\em 88}, 20011.
%
\bibitem{Refined} Rivas, A. Refined weak-coupling limit: Coherence, entanglement, and non-Markovianity. {\em  Phys. Rev. A} {\bf 2017}, {\em 95}, 042104.
%
\bibitem{Merkli} Merkli, M.; K\"onenberg, M. Completely positive dynamical semigroups and quantum resonance theory. {\em Lett. Math. Phys.} {\bf 2017}, {\em 107}, 1215.
%
\bibitem{DaviesSpohn}
Davies, E. B.; Spohn, H. Open quantum systems with time-dependent Hamiltonians and their linear response. {\em J. Stat. Phys.} {\bf 1978}, {\em 19}, 511.
%
\bibitem{noCP1} Benatti, F.; Floreanini, R.; Piani, M. Nonpositive evolutions in open system dynamics. {\em Phys. Rev. A} {\bf 2003}, {\em 67}, 042110.
%
\bibitem{noCP2} Benatti, F.; Floreanini, R.; Breteaux, S. Slipped nonpositive reduced dynamics and entanglement. {\em Laser Phys.} {\bf 2006}, {\em 16}, 1395.
%
\bibitem{noCP3} Anderloni, S.; Benatti, F.; Floreanini, R. Redfield reduced dynamics and entanglement. {\em J. Phys. A: Math.
Theor.} {\bf 2007}, {\em 40}, 1625.
%
\bibitem{Schaller2} Schaller, G.; Zedler, P.; Brandes, T. Systematic perturbation theory for dynamical coarse-graining {\em Phys. Rev. A} {\bf (2009)}, {\em 79}, 032110.
%
\bibitem{SWformula} Snider, R. F. Perturbation Variation Methods for a Quantum Boltzmann Equation. {\em J. Math. Phys.} {\bf (1964)}, {\em 5}, 1580. Wilcox, R. M. Exponential Operators and Parameter Differentiation in Quantum Physics. {\em J. Math. Phys.} {\bf (1967)}, {\em 8}, 962.
%
\bibitem{noMarkov1} Rivas, A.; Huelga, S. F.; Plenio, M. B. Quantum non-Markovianity: characterization, quantification and detection. {\em Rep. Prog. Phys.} {\bf 2014}, {\em 77},
094001.
%
\bibitem{noMarkov2} Breuer, H.-P.; Laine, E.-M.; Piilo, J.; Vacchini, B. Colloquium: Non-Markovian dynamics in open quantum systems. {\em Rev. Mod. Phys.} {\bf 2016}, {\em 88}, 021002.
%
\bibitem{noMarkov3} de Vega, I.;  Alonso, D. Dynamics of non-Markovian open quantum systems. {\em Rev. Mod. Phys.} {\bf 2017}, {\em 89}, 015001.
%
\bibitem{Spohn2} Spohn, H. An algebraic condition for the approach to equilibrium of an open N-level system. {\em Lett. Math. Phys.} {\em 1977}, {\em 2}, 33.
%
\bibitem{Wolf} Wolf, M., {\em Quantum channels \& operations: Guided tour} 2012, available at https://www-m5.ma.tum.de/foswiki/pub/M5/Allgemeines/MichaelWolf/QChannelLecture.pdf.
%
\bibitem{Monotonic} Lindblad, G. Completely positive maps and entropy inequalities. {\em Commun. Math. Phys.} {\bf 1975}, {\em 40} 147. Uhlmann, A. Relative entropy and the Wigner-Yanase-Dyson-Lieb concavity in an interpolation theory. {\em ibid.} {\bf 1977}, {\em 54} 21.
%
\bibitem{PMonotonic}  M\"uller-Hermes, A.; Reeb, D. Monotonicity of the Quantum Relative Entropy Under Positive Maps. {\em Ann. Henri Poincar\'e} {\bf 2017}, {\em 18}, 1777.
%
\bibitem{Wilde} Das, S.; Khatri, S.; Siopsis, G.;  Wilde, M.~M. Fundamental limits on quantum dynamics based on entropy change. {\em J. Math. Phys.} {\bf 2018}, {\em 59}, 012205.
%
\bibitem{MME} Rivas, A.; Plato, A.~D.~K.; Huelga, S. F.; Plenio, M.~B. Markovian master equations: a critical study. {\em New J. Phys.} {\bf 2010}, {\em 12}, 113032.
%
\bibitem{Puri} Puri, R.~R. {\em Mathematical Methods of Quantum Optics}; Springer: Berlin, Germany, 2001.


\end{thebibliography}
\end{document}